\documentclass{PoS}

\usepackage{amsmath}
\usepackage{graphicx}
\usepackage{epstopdf}
\usepackage{bm} 

\title{Large-volume results in SU(2) with adjoint fermions}

\ShortTitle{Large-volume SU(2) adjoint}

\author{\speaker{L. Del Debbio}\\
  The Higgs Centre for Theoretical Physics\\
  The University of Edinburgh, Edinburgh, UK\\
  E-mail: \email{Luigi.Del.Debbio@ed.ac.uk}}

\author{B. Lucini \\
  College of Science\\
  Swansea University, Swansea, UK \\
  E-mail: \email{b.lucini@swansea.ac.uk}} 

\author{C. Pica \\
  CP$^3$-Origins \& the Danish IAS, \\
  University of Southern Denmark, Odense, Denmark\\
  E-mail: \email{pica@cp3.dias.sdu.dk}} 

\author{A. Patella \\
  PH-TH, CERN, Geneva, Switzerland, and  \\
  School of Computing and Mathematics \& Centre for Mathematical
  Science,\\
  Plymouth University, Plymouth UK\\
  E-mail: \email{agostino.patella@plymouth.ac.uk}} 

\author{A. Rago \\
  School of Computing and Mathematics \& Centre for Mathematical
  Science,\\
  Plymouth University, Plymouth UK\\
  E-mail: \email{antonio.rago@plymouth.ac.uk}} 

\author{S. Roman\\
  The Higgs Centre for Theoretical Physics\\
  The University of Edinburgh, Edinburgh, UK\\
  E-mail: \email{S.Roman@ed.ac.uk}}

\abstract{Taming finite-volume effects is a crucial ingredient in
  order to identify the existence of IR fixed points. We present the
  latest results from our numerical simulations of SU(2) gauge theory
  with 2 Dirac fermions in the adjoint representation on large
  volumes. We compare with previous results, and extrapolate to
  thermodynamic limit when possible.}

\FullConference{31st International Symposium on Lattice Field Theory - LATTICE 2013\\
		July 29 - August 3, 2013\\
		Mainz, Germany}

\begin{document}

\section{Introduction}
\label{sec:intro}

After a decade of algorithmic and hardware improvements, light fermion masses can be simulated
effectively on large volumes using the current state-of-the-art computing resources.  As a
result current numerical techniques provide an effective tool to investigate the long-distance
dynamics of gauge theories from first principles, and several extensive studies have been
performed in recent years for vector gauge theories as the number of colors and flavors is
varied, and for fermionic matter in different representations of the color group -- see Kuti's
talk in these proceedings for a recent summary of results~\cite{kutilat13}. Besides their
academic interest, it would be particularly exciting to find a theory that could successfully
describe a strongly-interacting sector beyond the Standard Model (BSM).

A realistic candidate for BSM phenomenology must satisfy the constraints from electroweak
precision data, and lead to a successful construction of the SM flavor sector. It has been
clear for a long time that a rescaled version of QCD would not be adequate, and several
investigations have focussed on finding theories characterized by a large separation between
the IR and UV scales. Confining theories close to the edge of the conformal window are believed
to satisfy this requirement; the RG evolution of the couplings is very slow ({\em walking}),
which in turn leads to an approximate scale invariance at large distances. Gauge theories with
matter in higher representations could lead to this type of behaviour as argued in
Ref.~\cite{Hong:2004td}. Unfortunately a sharp characterization of walking behaviour has proved
elusive so far, despite some encouraging numerical evidence, see
e.g. Refs.~\cite{Fodor:2012ty,Appelquist:2012sm,Ishikawa:2013tua} for the latest papers, and
citations therein.

Proving the existence of an infrared fixed point (IRFP) is a well-defined question, for which
we can hope to find an unambiguous answer. Phenomenologically relevant models could then be
defined by deformations of the conformal theory, as suggested in Ref.~\cite{Luty:2004ye}. The
low-energy dynamics is then characterized by the anomalous dimensions of the relevant operators
at the putative IRFP. It has been known since the pioneering work in
Ref.~\cite{Yamawaki:1985zg} that large anomalous dimensions are required in order to construct
phenomenologically viable models. There are interesting bounds between these anomalous
dimensions that should be studied in more detail~\cite{Rattazzi:2008pe}. Finally it is
worthwhile to recall that realistic models have been constructed recently from a holographic
approach~\cite{Elander:2012fk}.  These studies should be complementary to the lattice ones presented in
this work.

The work reported here is part of our ongoing investigation of the existence of an IRFP in the
SU(2) gauge theory with two Dirac fermions in the adjoint representation, which is also known
under the name of Minimal Walking Technicolor (MWT)~\cite{Dietrich:2005jn}. This model is
believed to have a fixed point~\cite{Hietanen:2009az,Bursa:2009we,DelDebbio:2010hx}, and we
would like to obtain more quantitative results in particular about the mass anomalous dimension
at the IRFP. The systematic errors of lattice simulations need to be kept under control in
order to highlight the interesting conformal behaviour. Previous studies, which have been
useful in the early stages of the numerical investigations, are clearly affected by large
systematics and are useless in order to understand the physics of these theories.

In this note, we concentrate on the study of the spectrum of MWT. The systematic errors due to
the finite temporal and spatial extents of the lattice have been analysed in previous
publications, see e.g. Ref.~\cite{Bursa:2011ru}, where we found that systematic errors of
approximately 10\% are common on lattices such that $M_\mathrm{PS} L<10$, where $M_\mathrm{PS}$
indicates the mass of the lightest pseudoscalar state in the meson spectrum. We have therefore
embarked in large volume simulations of the theory, in order to obtain results for the
spectrum in a regime where systematic errors are below 1\%. 

The new volumes simulated are large enough to avoid finite temperature effects, and to allow an
extrapolation of the spectrum to the infinite volume limit for two values of the fermion
mass. These are the first results for the spectrum of the MWT that can be extrapolated to the
thermodynamical limit with an uncertainty at the percent level. The results for the full spectrum,
including glueball states, and the string tension, are discussed in Sect.~\ref{sec:results}.

\section{Methodology}
\label{sec:method}

We have simulated MWT using the RHMC algorithm described in Ref.~\cite{DelDebbio:2008zf}. In
order to identify, and control, the systematic effects due to the finite size of the lattices,
we have simulated the theory on a series of lattices, increasing both the temporal and the
spatial extent of the system. All simulations have been performed at fixed lattice bare
coupling $\beta=2.25$, and for two values of the fermion bare mass $am_0=-1.05,-1.15$. We have
also compared the spectrum obtained from simulations with the usual periodic boundary
conditions, to the one obtained with twisted boundary conditions, as defined in
Ref.~\cite{'tHooft:1979uj}. The details of our implementation will be presented in a
forthcoming publication. In the infinite volume limit, results should be independent of the
bounday conditions, and therefore we can use the dependence on the boundary conditions to
monitor whether or not the theory has reached the large volume asymptotic behaviour. The new
runs are listed in Tab.~\ref{tab:listruns}.

\begin{table}[ht]
  \centering
  \begin{tabular}{ccccccccc}
    \hline
    lattice & V & $-am_0$ & $N_\mathrm{traj}$ & $t_\mathrm{traj}$ & $\langle P\rangle$ & $\tau$ & $\lambda$ & $\tau_\lambda$ \\
    \hline
    A10 & $64\times 8^3$  & 1.15 & 810 & 3   & 0.66536(22) & 3.6(1.2) & 0.2005(58) & 1.42(31) \\
    A11 & $64\times 12^3$ & 1.15 & 530 & 1.5 & 0.66601(15) & 1.93(59) & 0.2054(40) & 1.54(43) \\
    C5 & $64\times 16^3$ & 1.15 & 1500 & 1.5 & 0.665992(61) & 2.32(46) & 0.2116(16) & 2.38(48) \\
    D4 & $64\times 24^3$ & 1.15 & 2387 & 1.5 & 0.665927(26) & 3.92(79) & 0.21478(70) & 1.70(24) \\
    F1 & $64\times 32^3$ & 1.15 & 2541 & 1.5 & 0.665946(30) & 3.37(62) & 0.2115(12) & 1.06(12) \\
    G1 & $80\times 48^3$ & 1.15 & 2200 & 1.5 & 0.665943(17) & 5.1(1.2) & 0.2237(17) & 0.637(58) \\
    \hline
    B2 & $24\times 12^3$ & 1.05 & 7819 & 1 & 0.647633(70) & 6.79(99) & 1.4936(51) & 5.80(78) \\
    C6 & $64\times 16^3$ & 1.05 & 2648 & 1.5 & 0.647645(48) & 4.63(96) & 1.4389(36) & 1.26(14) \\
    D5 & $64\times 24^3$ & 1.05 & 4000 & 1.5 & 0.647695(37) & 3.56(53) & 1.3906(45) & 0.722(54) \\
    F2 & $48\times 32^3$ & 1.05 & 3590 & 1.5 & 0.647680(30) & 4.28(74) & 1.3708(49) & 0.632(45) \\
    \hline
    \hline
    TWA1 & $64\times 8^3$  & 1.15 & 565 & 1.5 & 0.66665(22) & 2.8(1.0) & 0.5557(96) & 0.85(18) \\
    TWB1 & $64\times 12^3$ & 1.15 & 741 & 1.5 & 0.66590(11) & 2.96(96) & 0.2709(48) & 1.93(50) \\
    TWC1 & $64\times 16^3$ & 1.15 & 1162 & 1.5 & 0.665990(61) & 2.91(73) & 0.2484(17) & 6.2(2.2) \\
    TWD1 & $64\times 24^3$ & 1.15 & 2701 & 1.5 & 0.665912(35) & 4.63(95) & 0.21840(88) & 2.43(37) \\
    TWE1 & $160\times 16^3$ & 1.217468 & 710 & 2 & -- & -- & -- & -- \\
    \hline
  \end{tabular}
  \caption{List of lattices used in this study. $\langle P\rangle$ is
    the average value of the plaquette, $\tau$ is its integrated autocorrelation
    time; $\lambda$ is the lowest eigenvalue of the Hermitian Dirac
    matrix squared $H^2$, and again $\tau_\lambda$ is the
    corresponding autocorrelation time. The bare mass for the TWE1
    lattice is tuned to the chiral limit. }
  \label{tab:listruns}
\end{table}

\subsection{Observables}
\label{sec:observables}

Mesonic observables are extracted from two-point functions of fermion bilinears:
\begin{equation}
  \label{eq:mes2pt}
  f_{\Gamma\Gamma^\prime}(t) = \sum_{\vec{x}}\, \langle \Phi_\Gamma(\vec{x},t)^\dagger
  \Phi_{\Gamma^\prime}(\vec{0},0)\rangle\, , 
\end{equation}
where 
\begin{equation}
  \label{eq:fbil}
  \Phi_\Gamma(\vec{x},t) = \bar\psi_1(\vec{x},t) \Gamma \psi_2(\vec{x},t)\, .
\end{equation}
Note that we always consider non-singlet flavor states, as indicated by the indices $1,2$ that
appear in the definition of the fermion bilinear. The matrices $\Gamma$ and $\Gamma^\prime$ act in
spin space, and determine the quantum numbers of the states that contribute to the correlator in
Eq.~(\ref{eq:mes2pt}). The details of the analysis used to extract the PCAC mass, the hadron
masses, and the decay constants are explained in detail in the Appendix of
Ref.~\cite{Bursa:2011ru}.

Gluonic observables are obtained using a variational method and a large basis of operators, as described in
Ref.~\cite{Lucini:2010nv}.

\subsection{Finite temperature effects}
\label{sec:finite-temp-effects}

\begin{figure}[ht] 
  \includegraphics*[scale=0.35]{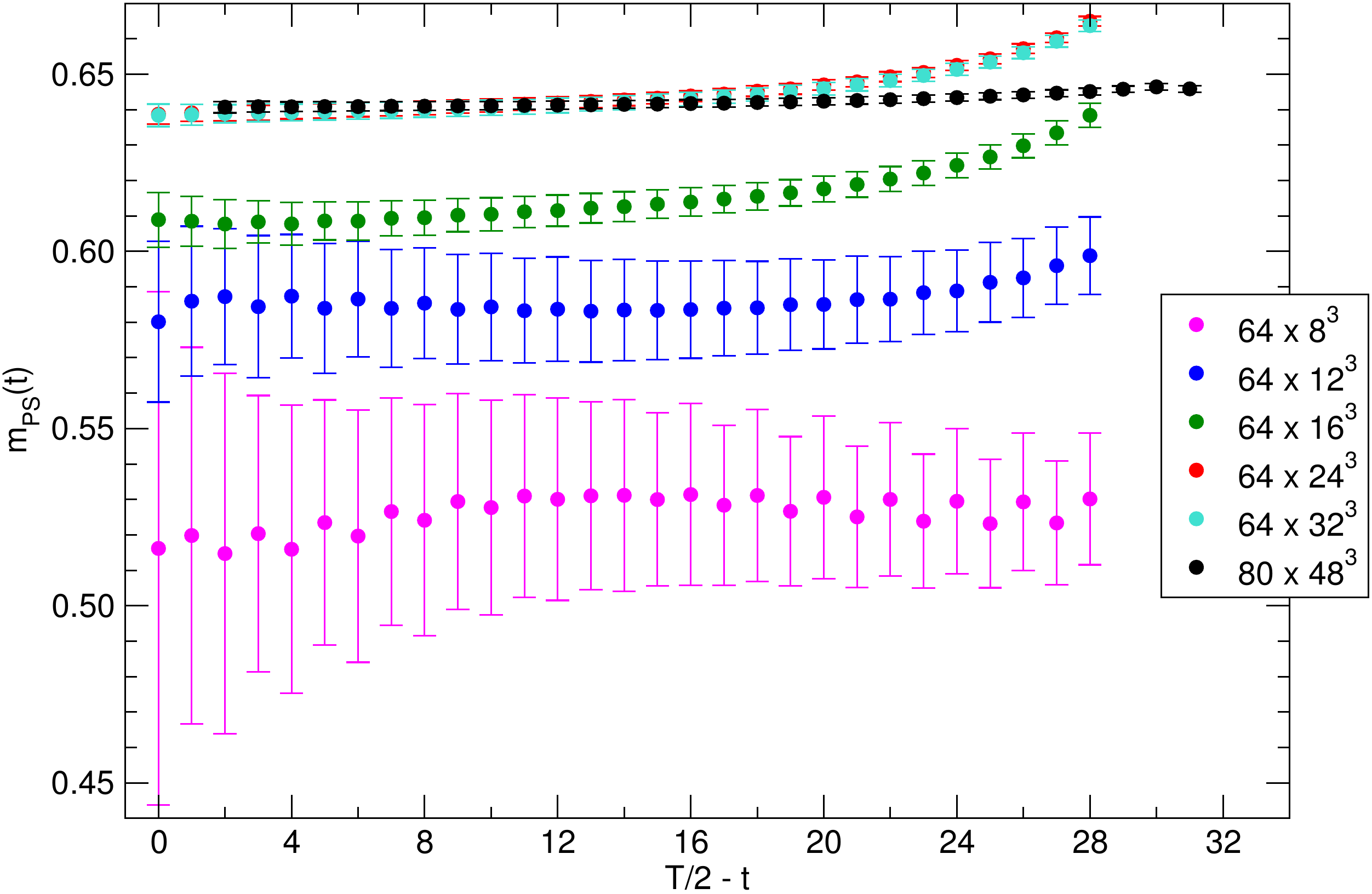} 
  \caption{Plateaux for the effective mass in the PS channel at $\beta=2.25$, and $am_0=-1.15$. It
    is clear from the figure that the plateaux as a function of $t$ are clearly identified for the
    current range of parameters, and that there is no contamination from higher states in the
    spectrum for the lattices considered here, ie with $T/a \geq 64$.}
\label{fig:15plateaux}
\end{figure}

We had noticed in our previous simulations that lattices with a temporal extent $T/a=16, 24$ are not
long enough to identify unambiguously the onset of the asymptotic behaviour of the field
correlators~\cite{Bursa:2011ru}. Larger time extensions have been used in this study in order to be
able to identify long plateaux in the two-point correlators, and to fit the large time behaviour in
a regime where contaminations from higher states in the spectrum are suppressed. A typical set of
plateaux for the mass of the pseudoscalar meson is reported in Fig.~\ref{fig:15plateaux}. It is clear
from the plot, that lattices with temporal extent $T/a\geq 64$ exhibit very clear plateaux, which
can be easily fitted to a constant over a large range in $t$. As a consequence, we decided not to
use smeared sources for this analysis, since the introduction of smeared sources increases the
auto-correlation time of the spectral observables.

\section{Results}
\label{sec:results}

\begin{figure}[ht] 
  \includegraphics*[scale=0.35]{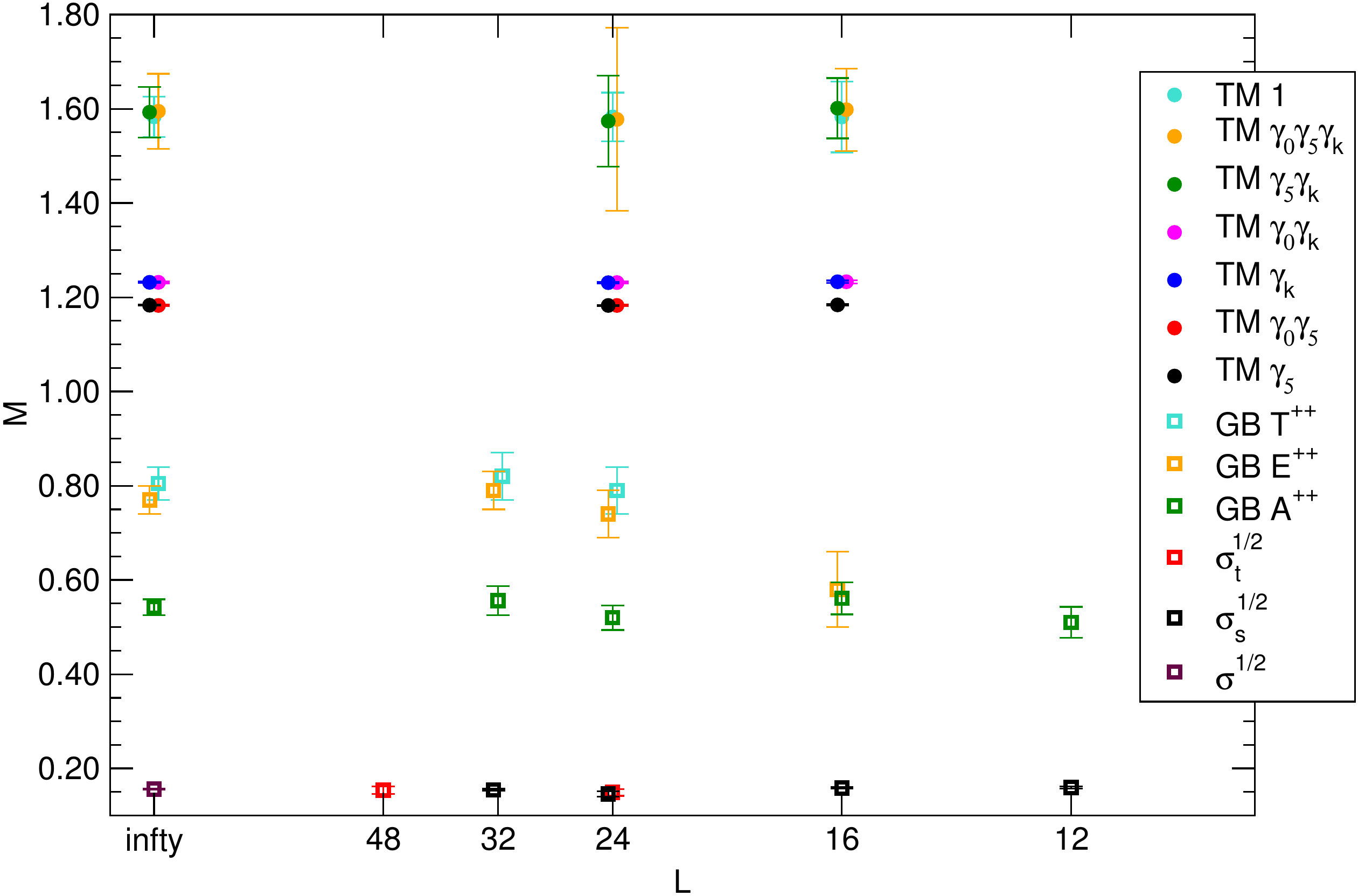} 
  \caption{ Summary of the spectrum for all the states included in this analysis at
    $am_0=-1.05$. The data are shown as a function of the number of sites in the spatial
    directions $L/a$.  }
\label{fig:05all}
\end{figure}

The heaviest fermion mass in the new set of simulations is $am_0=-1.05$, which corresponds to a
PCAC mass of $am=0.2688(15)$. The full spectrum is shown in Fig.~\ref{fig:05all}. As observed in
previous studies the mesons are heavier than the glueballs, with the smallest scale being set by the
string tension.
At this value of the fermion mass, the finite volume effects on the mesonic states are small,
and we can confidently extrapolate their values to the thermodynamical limit. Different
combinations of $\Gamma,\Gamma^\prime$ that project onto the same physical states yield results
that are compatible within statistical errors. The new simulations confirm the results that we
presented in previous studies~\cite{Bursa:2011ru}.  For the string tension we see a discrepancy
between the value of the string tension computed on the smaller lattices, and the one obtained
on the larger volumes. Simulations on sufficiently large lattices, $L/a\geq 24$ yield consistent
results between the spatial and temporal string tensions.

Finite volume effects are more significant for the lighter fermion mass. It is clear from
Fig.~\ref{fig:15plateaux} that the mass of the pseudoscalar is affected by sizeable finite
volume corrections on the smaller lattices. The fitted value converges for lattices with
$L/a\geq 24$. It is clear from this plot, that the masses are affected by large finite-volume
effects on the smaller lattices, and that the effective mass is growing as the spatial size of
the lattice increased. This is consistent with the observation that, in a theory where chiral
symmetry is not spontaneously broken, the pseudoscalar mass reaches the infinite volume limit
from below, as explained in Ref.~\cite{Patella:2011kp,Luscher:1985dn}.

The plateaux in the effective mass of the pseudoscalar also show that a large temporal extent
is needed in order to suppress the corrections from heavier states in the same channel. This is
expected for a mass deformed conformal theory, since {\em all} the states in the spectrum
should scale to zero following the same power-law scaling formula. The contribution from
excited states becomes neglibile when: 
\begin{equation}
  \label{eq:Deltam}
  t \Delta M_\mathrm{PS} \gg 1\, 
\end{equation}
where $\Delta M_\mathrm{PS}$ is the mass difference between the ground state and the first
excited stated in the pseudoscalar channel. For a conformal theory near the chiral limit,
$\Delta M_\mathrm{PS} \propto M_\mathrm{PS}$, so that the exponential suppression of the
excited states becomes really effective only at large temporal separations.

\begin{figure}[ht] 
  \includegraphics*[scale=0.35]{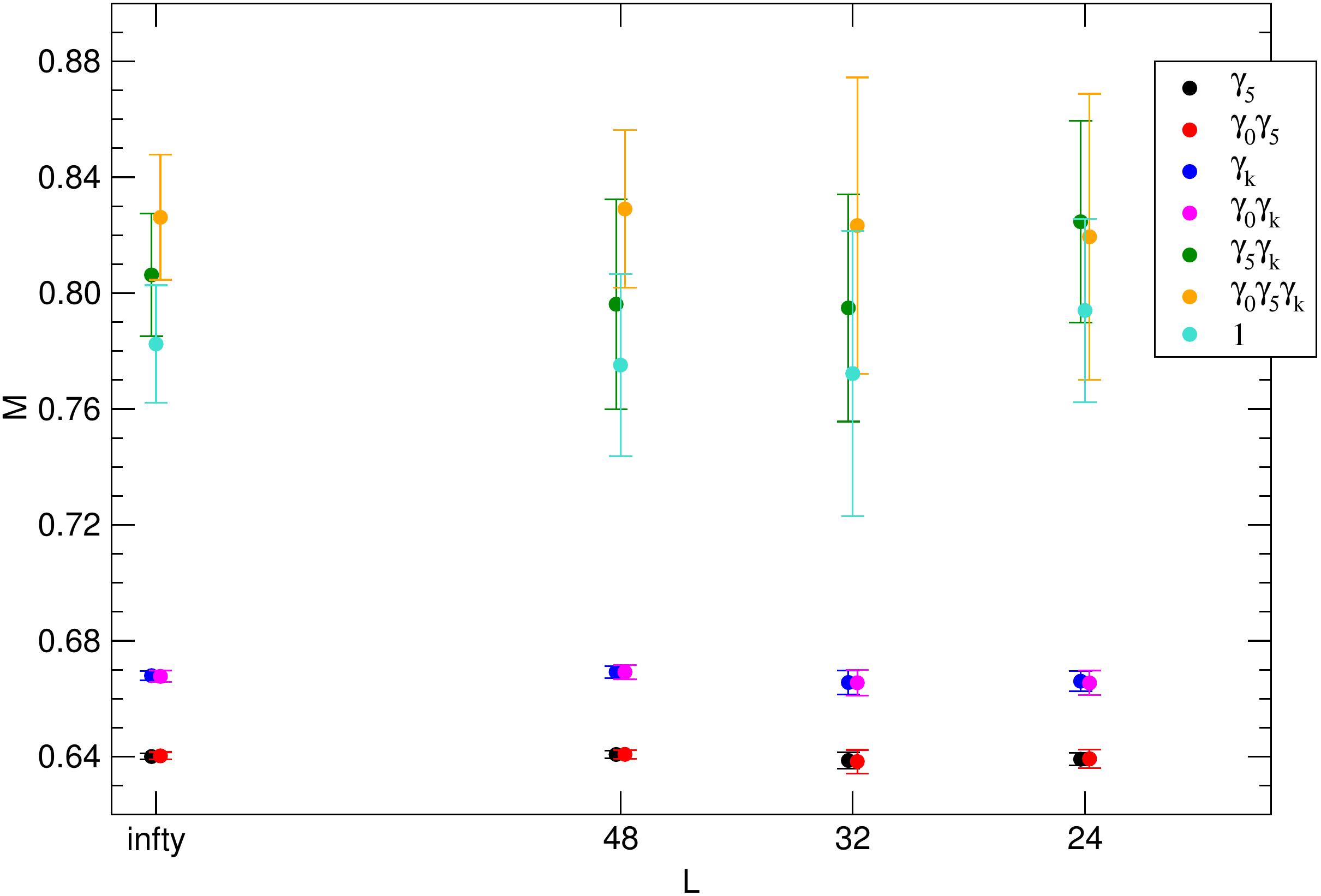} 
  \caption{ Summary of the spectrum of the mesonic states included in this analysis at
    $am_0=-1.15$. The data are shown as a function of the number of sites in the spatial
    directions $L/a$.  }
\label{fig:15mesons}
\end{figure}

Data for the mesonic spectrum at $am_0=-1.15$ are displayed in Fig.~\ref{fig:15mesons}, showing
that a reliable extrapolation to the infinite-volume limit is possible using data for lattices
with $L/a\geq 24$. This conclusion is supported by the observed independence of the mesonic
spectrum form the choice of boundary conditions for $L/a \geq
24$. Note that for
this choice of the bare fermion mass, the PCAC mass is $am= 0.11792(63)$.

\begin{figure}[ht] 
  \includegraphics*[scale=0.35]{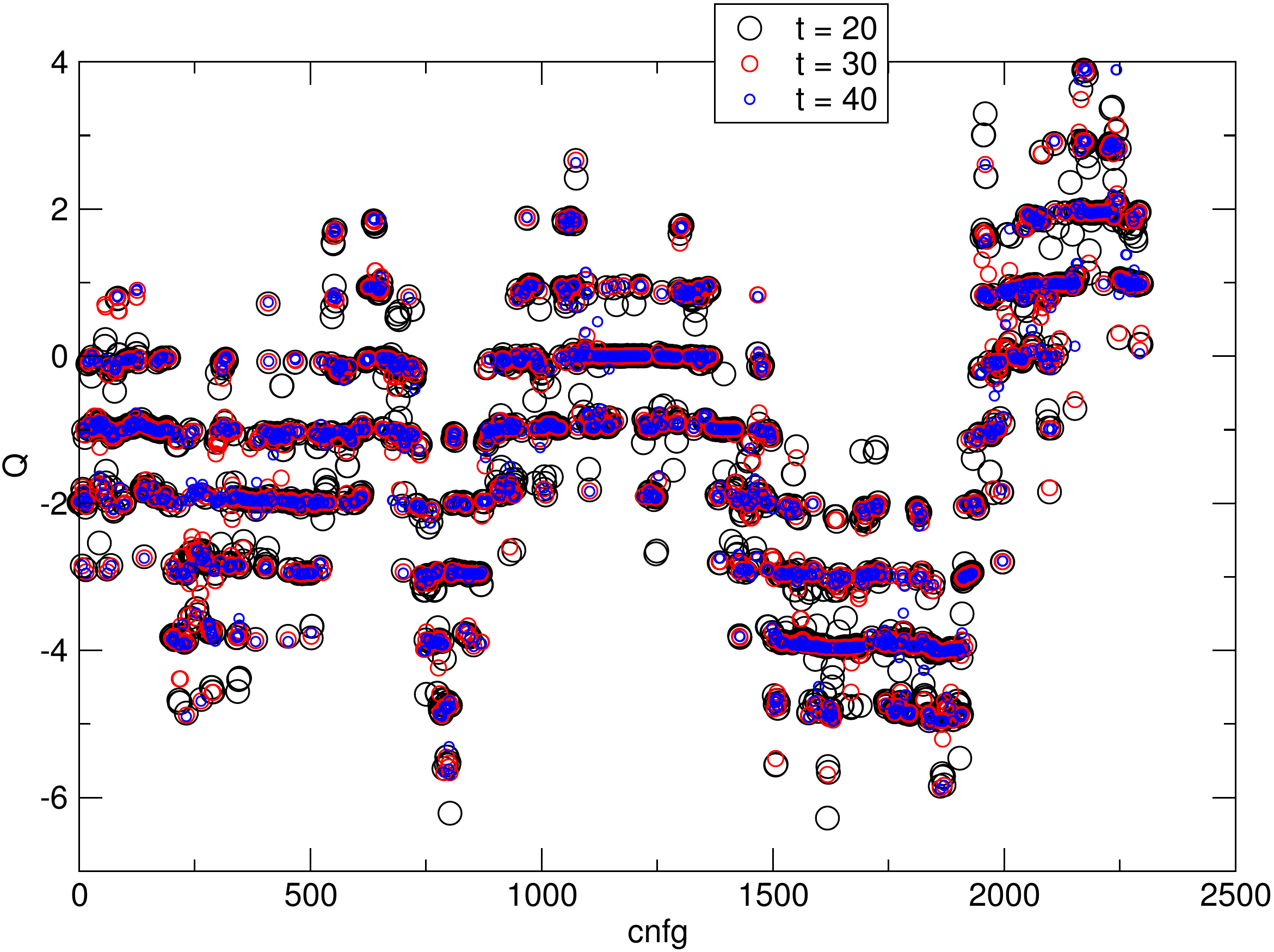} 
  \caption{
    Time history for the topological charge as a function of the configuration number along the
    HMC evolution. Values for different values of the Wilson flow time are reported. 
  }
\label{fig:topo2}
\end{figure}

The time-history of the topological charge along the HMC evolution is shown in
Fig.~\ref{fig:topo2}. The topological charge is computed using a bosonic estimator for
different values of the Wilson flow time $t$. With parameters currently in use, the HMC seems
to sample the topology correctly. A more detailed investigation of these issues is required. 

\begin{figure}[ht] 
  \includegraphics*[scale=0.25]{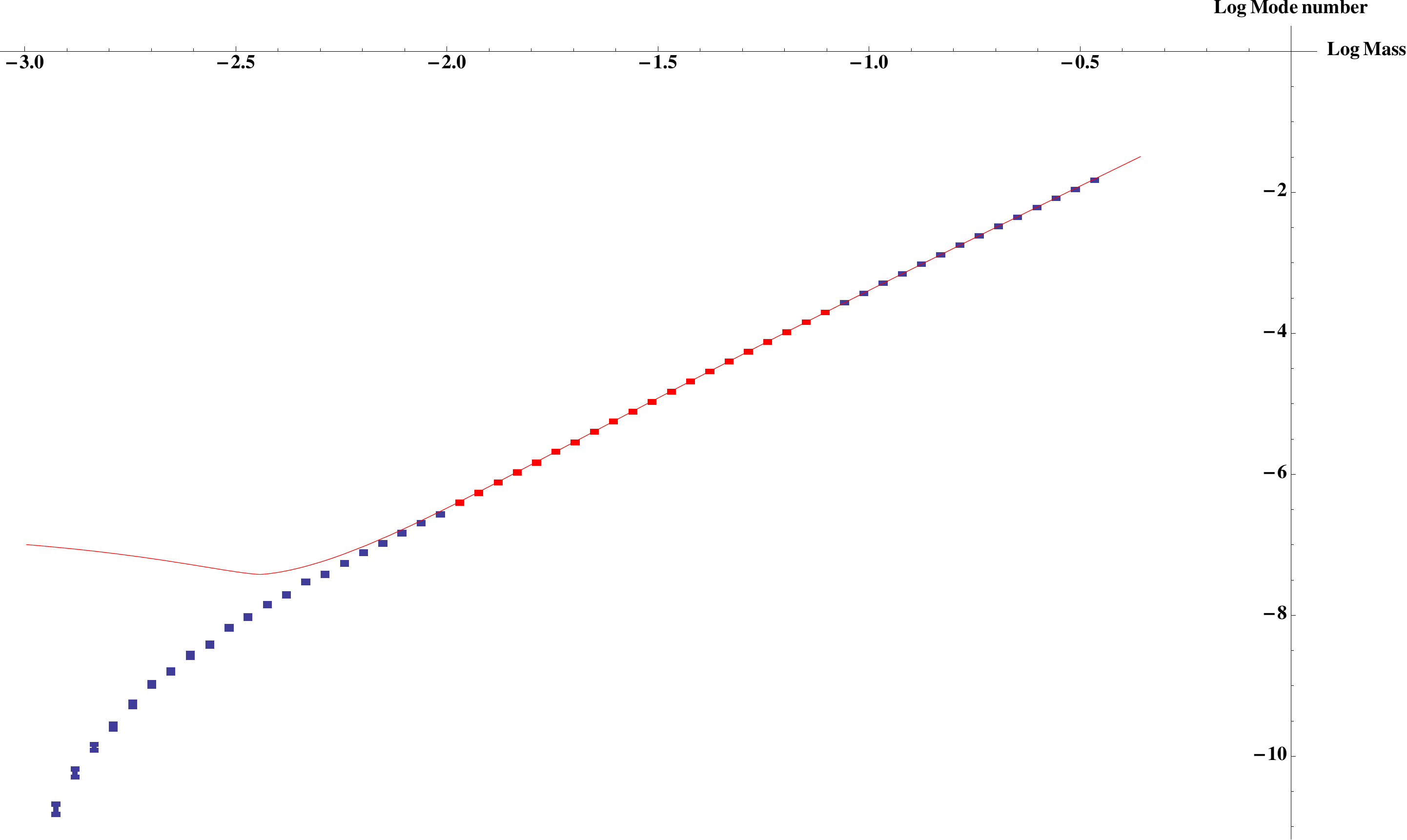} 
  \caption{
    Mode number for the Dirac operator as a function of the  measured on the lattice TWE1. The
    curve shows a fit to the predicted scaling behaviour. These results are preliminary.
  }
\label{fig:eigden}
\end{figure}

The density of eigenvalues of the Dirac operator has proved to be a valuable tool to measure
the mass anomalous dimension in a conformal
theory~\cite{DeGrand:2009hu,DelDebbio:2010ze,Patella:2012da}. Using the techniques implemented
in Ref.~\cite{Patella:2012da}, we have measured the density of eigenvalues on the twisted
lattices. Denoting by $\rho(\omega)$ the eigenvalue density for the operator $(D+m)^\dagger
(D+m)$,  the mode number is defined as
\begin{equation}
  \label{eq:modenum}
  \nu(M) = 2 \int_0^{\sqrt{M^2-m^2}} \! d\omega\, \rho(\omega)\, .
\end{equation}
The expected scaling is dictated by the mass anomalous dimension: 
\begin{equation}
  \label{eq:modescal}
  a^{-4} \nu(M) = a^{-4} \nu_0 + A \left[ (aM)^2 - (am)^2\right]^{\frac{2}{1+\gamma^*}}\,.
\end{equation}
The results for the lattice TWE1 are shown in Fig.~\ref{fig:eigden}. As expected on
theoretical grounds, an intermediate
region in the value of the mode number can be identified where a power-law yields an excellent
fit to the data. The fit yields a determination of the mass anamalous dimension 
\begin{equation}
  \label{eq:andimfit}
  \gamma_* = 0.38(2)\, ,
\end{equation}
in agreement with the value obtained in Ref.~\cite{Patella:2012da}. These preliminary results
are encouraging, and will be finalised in a forthcoming work. 

\section{Outlook}
\label{sec:outlook}

Our latest simulations show convincingly that finite volume effects can be kept under control in
the spectrum of MWT. The analysis for the glueball spectrum is being finalised, and we expect
to have a determination of the whole spectrum with controlled systematics. We have started to
monitor the topology, in order to verify that our simulations are not stuck at fixed topology. 
Last but not least, we have looked at the eigenvalue density, and computed the mass anomalous
dimension on one of the lattices with twisted boundary conditions. The twisted boundary
conditions allow us to simulate at very small fermion mass, and our analysis confirms the
results obtained on periodic lattices in earlier studies. 

The preliminary results reported here are in agreement with our previous findings for this
model, a comprehensive report of these results is in preparation.


\begin{thebibliography}{99}
\bibitem{kutilat13} J.~Kuti, plenary tak, these proceedings. 

\bibitem{Hong:2004td} 
  D.~K.~Hong, S.~D.~H.~Hsu and F.~Sannino,
  Phys.\ Lett.\ B {\bf 597}, 89 (2004)
  [hep-ph/0406200].

\bibitem{Fodor:2012ty} 
  Z.~Fodor, K.~Holland, J.~Kuti, D.~Nogradi, C.~Schroeder and C.~H.~Wong,
  Phys.\ Lett.\ B {\bf 718}, 657 (2012)
  [arXiv:1209.0391 [hep-lat]].

\bibitem{Appelquist:2012sm} 
  T.~Appelquist, R.~Babich, R.~C.~Brower, M.~I.~Buchoff, M.~Cheng, M.~A.~Clark, S.~D.~Cohen and G.~T.~Fleming {\it et al.},
  Phys.\ Rev.\ D {\bf 85}, 074505 (2012)
  [arXiv:1201.3977 [hep-lat]].

\bibitem{Ishikawa:2013tua} 
  K.~-I.~Ishikawa, Y.~Iwasaki, Y.~Nakayama and T.~Yoshie,
  arXiv:1310.5049 [hep-lat].

\bibitem{Luty:2004ye} 
  M.~A.~Luty and T.~Okui,
  JHEP {\bf 0609}, 070 (2006)
  [hep-ph/0409274].

\bibitem{Yamawaki:1985zg} 
  K.~Yamawaki, M.~Bando and K.~-i.~Matumoto,
  Phys.\ Rev.\ Lett.\  {\bf 56}, 1335 (1986).

\bibitem{Rattazzi:2008pe} 
  R.~Rattazzi, V.~S.~Rychkov, E.~Tonni and A.~Vichi,
  JHEP {\bf 0812}, 031 (2008)
  [arXiv:0807.0004 [hep-th]].

\bibitem{Elander:2012fk} 
  D.~Elander and M.~Piai,
  Nucl.\ Phys.\ B {\bf 867}, 779 (2013)
  [arXiv:1208.0546 [hep-ph]].

\bibitem{Dietrich:2005jn} 
  D.~D.~Dietrich, F.~Sannino and K.~Tuominen,
  Phys.\ Rev.\ D {\bf 72}, 055001 (2005)
  [hep-ph/0505059].

\bibitem{DelDebbio:2010hx} 
  L.~Del Debbio, B.~Lucini, A.~Patella, C.~Pica and A.~Rago,
  Phys.\ Rev.\ D {\bf 82}, 014510 (2010)
  [arXiv:1004.3206 [hep-lat]].

\bibitem{Hietanen:2009az} 
  A.~J.~Hietanen, K.~Rummukainen and K.~Tuominen,
  Phys.\ Rev.\ D {\bf 80}, 094504 (2009)
  [arXiv:0904.0864 [hep-lat]].

\bibitem{Bursa:2009we} 
  F.~Bursa, L.~Del Debbio, L.~Keegan, C.~Pica and T.~Pickup,
  Phys.\ Rev.\ D {\bf 81}, 014505 (2010)
  [arXiv:0910.4535 [hep-ph]].

\bibitem{Bursa:2011ru} 
  F.~Bursa, L.~Del Debbio, D.~Henty, E.~Kerrane, B.~Lucini, A.~Patella, C.~Pica and T.~Pickup {\it et al.},
  Phys.\ Rev.\ D {\bf 84}, 034506 (2011)
  [arXiv:1104.4301 [hep-lat]].

\bibitem{DelDebbio:2008zf} 
  L.~Del Debbio, A.~Patella and C.~Pica,
  Phys.\ Rev.\ D {\bf 81}, 094503 (2010)
  [arXiv:0805.2058 [hep-lat]].

\bibitem{'tHooft:1979uj} 
  G.~'t Hooft,
  Nucl.\ Phys.\ B {\bf 153}, 141 (1979).

\bibitem{Lucini:2010nv} 
  B.~Lucini, A.~Rago and E.~Rinaldi,
  JHEP {\bf 1008}, 119 (2010)
  [arXiv:1007.3879 [hep-lat]].

\bibitem{Patella:2011kp} 
  A.~Patella, L.~Del Debbio, B.~Lucini, C.~Pica and A.~Rago,
  PoS LATTICE {\bf 2011}, 084 (2011)
  [arXiv:1111.4672 [hep-lat]].

\bibitem{Luscher:1985dn} 
  M.~Luscher,
  Commun.\ Math.\ Phys.\  {\bf 104}, 177 (1986).

\bibitem{DeGrand:2009hu} 
  T.~DeGrand,
  Phys.\ Rev.\ D {\bf 80}, 114507 (2009)
  [arXiv:0910.3072 [hep-lat]].

\bibitem{DelDebbio:2010ze} 
  L.~Del Debbio and R.~Zwicky,
  Phys.\ Rev.\ D {\bf 82}, 014502 (2010)
  [arXiv:1005.2371 [hep-ph]].

\bibitem{Patella:2012da} 
  A.~Patella,
  Phys.\ Rev.\ D {\bf 86}, 025006 (2012)
  [arXiv:1204.4432 [hep-lat]].

\end{thebibliography}
\end{document}